\documentclass[%
 aip,
 amsmath,amssymb,
 reprint,%
]{revtex4-1}

\usepackage{graphicx}
\usepackage{dcolumn}
\usepackage{bm}

\usepackage[utf8]{inputenc}
\usepackage[T1]{fontenc}
\usepackage{mathptmx}
\usepackage{etoolbox}
\usepackage{tabularx}
\usepackage{multirow}
\usepackage{hyperref}
\usepackage[noabbrev]{cleveref}
\usepackage[version=3]{mhchem}
\usepackage[text-series-to-math=true]{siunitx}

\makeatletter
\def\@email#1#2{%
 \endgroup
 \patchcmd{\titleblock@produce}
  {\frontmatter@RRAPformat}
  {\frontmatter@RRAPformat{\produce@RRAP{*#1\href{mailto:#2}{#2}}}\frontmatter@RRAPformat}
  {}{}
}%
\makeatother
\begin{document}

\preprint{AIP/123-QED}

\title[]{Photo-induced non-thermal lattice disorder in aluminium thin-film}
\author{F. Rodriguez Diaz}
\affiliation{Max-Born-Institut, Max-Born-Stra\ss e 2A, 12489 Berlin, Germany}
\affiliation{Current address: Physics Department, Universit\"at Hamburg, Germany.}
\author{K. Amini}%
\email{kasra.amini@mbi-berlin.de}
\affiliation{Max-Born-Institut, Max-Born-Stra\ss e 2A, 12489 Berlin, Germany}


\date{\today}

\begin{abstract}
We investigate the ultrafast dynamics of photo-induced non-thermal lattice disorder in a polycrystalline aluminium thin film to elucidate transient short- and long-range lattice distortions, their thermalization and electron-phonon coupling timescales. Using a high-repetition-rate 95-keV ultrafast electron diffraction setup (UED), we measured the transient dynamics for the differential scattering signal in a momentum transfer, $q$, range longer than in conventional keV UED setups. Analysis of ten Bragg and six diffuse scattering revealed a prompt increase in the mean-square displacement (MSD), indicating rapid energy transfer from the excited electronic system to the lattice. The subsequent relaxation dynamics of the elastic scattering intensities exhibit a pronounced dependence on diffraction order. Lower-order reflections relax more rapidly, whereas higher-order reflections show significantly slower relaxation or near-plateau behaviour, indicating that lattice equilibration proceeds on multiple $q$-dependent timescales. Exponential fits to the MSD dynamics reveal oscillatory residuals, indicative of coherent non-thermal lattice motion. Power spectral density analysis of these residuals uncovers coherent lattice oscillations with a fundamental frequency of $\omega_0 = 0.192$ THz, corresponding to the acoustic breathing (A$_{1g}$) mode of aluminium. Higher frequency components are also observed, consistent with coherent phonon oscillations originating from a single zero-wavevector mode populated by multiple coherent phonons. While individual phonon branches are not directly resolved, the observed dependence on the lattice plane of the relaxation behaviour and oscillatory signatures are consistent with a mode-selective lattice response and non-thermal energy redistribution as described by non-thermal lattice models.
\end{abstract}

\maketitle

\section{Introduction}

\par The optical excitation of metals initiates a cascade of energy transfer processes, with the population of hot electrons that couple to the lattice on sub-picosecond timescales \cite{Waldecker2016,Mo2018,Zahn2021}. The following relaxation mechanisms are influenced by the excitation and redistribution of phonon populations, lasting several picoseconds. 

\par Ultrafast structural dynamics of metals undergoing phase transitions have been the subject of interest since the first UED studies by Williamson and Mourou \cite{Williamson1984,Mourou1982} demonstrating the ability to directly probe atomic-scale structural changes on picosecond timescales during laser-induced melting in aluminium. Later, Guo \emph{et al.} \cite{Guo2000} established a clear distinction between thermal and non-thermal laser-induced solid-to-liquid transitions in aluminium, identifying a fluence threshold of 34 \si{mJ/cm^2} for non-thermal melting. Their reflectivity measurements highlighted the crucial role of interband electronic excitation at 800 nm, as evidenced by changes in the dielectric response. Subsequently, Siwick \emph{et al.}\cite{Siwick2003} used femtosecond UED to study irreversible melting of aluminium at fluences well above the damage threshold, demonstrating that long-range crystalline order is lost within several picoseconds, while the short-range order persists on longer timescales. Despite these advances, the pathways governing non-thermal lattice disordering in metals, particularly below the melting threshold, remain an active area of research \cite{Zhang2022}. 

\par Using the two-temperature model \cite{Anisimov1974} within a Debye-Waller framework\cite{Gao1999}, Park \emph{et al.} \cite{Park2005a,Park2005} perform UED measurements demonstrating coherent atomic displacements in photoexcited aluminium using UED, at excitation fluences in the non-thermal regime. They identified coherent lattice motion associated with a longitudinal acoustic (LA) \emph{breathing} mode in polycrystalline thin films. Beyond Bragg scattering, diffuse scattering provides direct sensitivity to non-equilibrium phonon populations and lattice disorder beyond average unit-cell displacements\cite{Otto2021}. As originally proposed and simulated by Rahman \emph{et al.} \cite{Rahman1985}, the temporal evolution of diffuse scattering contains information about energy flow from hot electrons into the lattice and about phonon dynamics. Using time-resolved ultrafast electron diffuse scattering (UEDS), Zhu \emph{et al.} \cite{Zhu2013} demonstrated that, in addition to long-range lattice distortions such as coherent breathing modes, short-range lattice disorder develops on faster timescales. This suggests the presence of multiple lattice equilibration pathways.

\par Building on these experimental insights, Waldecker \emph{et al.} \cite{Waldecker2016} introduced a non-thermal lattice model (NLM) that extends the conventional two-temperature model \cite{Anisimov1974,Allen1987} by explicitly accounting for branch-dependent electron-phonon coupling and transient non-thermal phonon populations based on LA and transverse acoustic (TA) phonons. In this framework, different phonon subsystems equilibrate on distinct timescales, implying that lattice disorder and relaxation dynamics should exhibit a pronounced dependence on momentum transfer. Experimental validation of such models, therefore, requires measurements that resolve lattice dynamics in both time and reciprocal space, rather than relying on momentum-space-averaged observables.

\par In this work, we address this challenge by combining high-repetition-rate UED with high dynamic range and single-electron detection to investigate photo-induced non-thermal lattice disorder in a polycrystalline aluminium thin film below the damage threshold. By analysing Bragg and diffuse scattering signals separately for individual diffraction orders, we directly resolve momentum-dependent lattice dynamics, Bragg-planes-specific relaxation times, and coherent oscillatory responses. This approach provides further insights into non-thermal lattice dynamics and enables a detailed examination of energy redistribution processes in photoexcited aluminium by comparing and correlating the responses of Bragg-plane-specific diffraction signals.

\section{Experiments and Methods:}

\begin{figure*}
	\includegraphics[width=0.8\textwidth]{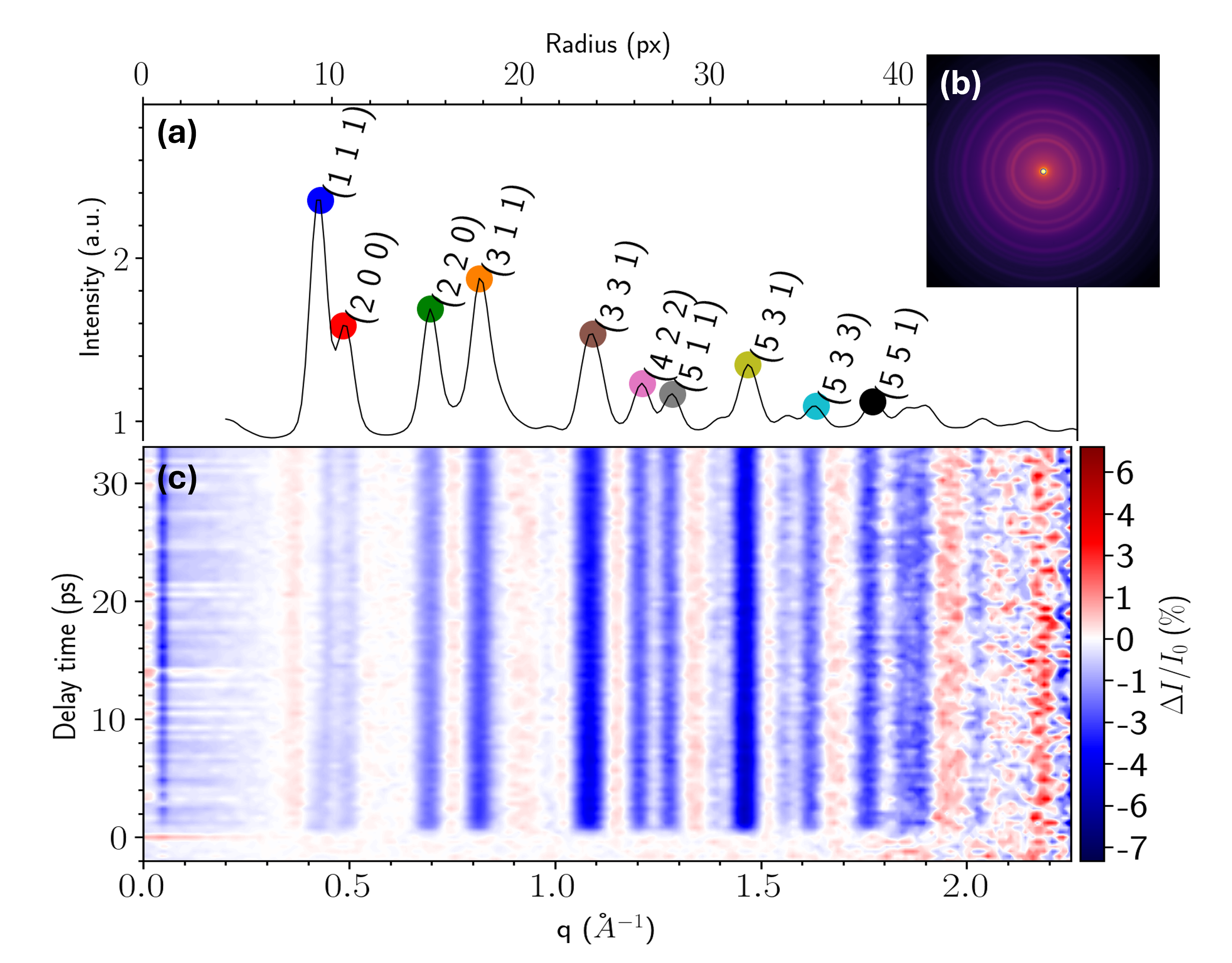}
	\caption{(a) Measured radial intensity distribution of the electron diffraction pattern from a 31-nm polycrystalline aluminium thin film, with Bragg reflections labelled by Miller indices. (b) Electron diffraction pattern measured at negative pump-probe delays using 95 keV electron pulses and a direct electron detector. The blue circle indicates the primary unscattered electron beam used for intensity normalization. (c) Measured two-dimensional map of the differential scattering signal, $\Delta I/I_0$, as a function of the reciprocal space, $q$, and pump-probe delay, $\Delta t$. \label{fig:aluminium_ediff}}
\end{figure*}

\subsection{Ultrafast electron diffraction setup}

\par Time-resolved measurements were performed using the high-repetition rate electron diffraction setup described in Ref.~\cite{Rodriguez-Diaz2024}. A polycrystalline aluminium thin film (Plano GmbH, S108, 31\,$\pm$\,2 nm thickness) supported on a copper TEM grid (400 lines/inch) was excited by an 800 nm pulse (60 fs, 2 \si{mJ/cm^2}, 200 \si{\micro m} FWHM beam diameter) and probed by a 95 kV temporally uncompressed electron pulses ($\sim$300 fs, 22 aC or 137 electrons per pulse, 100 \si{\micro m} FWHM beam diameter) at a repetition rate of 30 kHz. Pump-probe delays from -20 ps to +80 ps were scanned in 0.4 ps steps. More than 30 individual complete pump-probe delay scans were averaged, with a 60 s integration time (1.8 million shots) per delay, yielding statistics 30 ($\sim$100) times higher than those of typical keV (MeV) UED measurements. A 200 \si{\micro m} aperture placed in front of the aluminium sample maximized the photoexcited area while minimizing dark-current background and cumulative thermal heating \cite{Domrose2025}. Diffraction patterns were recorded using a direct electron detector (DECTRIS QUADRO) with 84\% detective quantum efficiency at 95 keV, enabling simultaneous detection of intense low-order reflections and weak high-order reflections\cite{Rodriguez-Diaz2024}. 

\subsection{Data analysis}

Radial intensity profiles were obtained by azimuthally averaging the two-dimensional diffraction patterns. The differential scattering signal, \[\Delta I/I_0 = \frac{I_{(t>T_0)} - I_{(t<T_0)}}{I_{(t<T_0)}},\] where $t$ is the pump-probe delay and $T_0$ is the pump-probe time zero, was analysed for ten Bragg scattering (BS) reflections and six diffuse scattering (DS) regions located between Bragg peaks. 

The temporal evolution was fitted using

\begin{align}
	f^-(t;\mathrm{a}) &= \left(\mathrm{a}_0 \, \mathrm{erfc}\left(
	\frac{t-2\sigma_{{\mathrm{decay}}}}{\sqrt{2}\sigma_{{\mathrm{decay}}}}
	\right) + \mathrm{a}_1\right) \exp{\left(t/\tau_r\right)}, \label{eq:ccdf_exp} \\
	f^+(t;\mathrm{a}) &= \left(\mathrm{a}_0 \, \mathrm{erf}\left(
	\frac{t-2\sigma_{{\mathrm{rise}}}}{\sqrt{2}\sigma_{{\mathrm{rise}}}}
	\right) + \mathrm{a}_1\right) \exp{\left(t/\tau_r\right)}, \label{eq:cdf_exp}
\end{align}

\noindent where $f^-$ and $f^+$ describe Bragg (decay) and diffuse (rise) responses. The error function captures the initial step-function intensity change due to electron-phonon coupling, while the exponential term accounts for the subsequent recovery of the signal. The parameters \(\rm{a}_0\) and \(\rm{a}_1\) represent amplitude values, while \(\sigma_{{\rm{decay}}}\) and \(\sigma_{{\rm{rise}}}\) represent the root-mean-square (RMS) values for the decay and rise times of the respective signals. Here, we report decay and rise times as $\tau_{\mathrm{decay}} = 
4\sigma_{{\mathrm{decay}}}$ and $\tau_{\mathrm{rise}} = 4\sigma_{{\mathrm{rise}}}$.

The mean-square displacement (MSD) was extracted from the relative changes in Bragg peak intensities using the Debye–Waller relation\cite{Waldecker2016} given by

\begin{equation}\label{eq:MSD}
	\Delta\langle u^2_{\mathrm n}(t)\rangle-\Delta\langle u^2_{0\mathrm n}(t)\rangle=\frac{3}{4}\left(\frac{1}{\pi q_{\mathrm n}}\right)^2\ln\left(\frac{I_{0\mathrm n}}{I_{\mathrm n}}\right),
\end{equation}

\noindent where $q_{\mathrm n}$ and $I_{\mathrm n}$ represent the momentum transfer and intensity of the $n^\mathrm{th}$ Bragg reflection, respectively, and $I_{0\mathrm n}$ is the probe-only reference intensity. This expression assumes an isotropic, thermally equilibrated displacement distribution, with deviations between MSD values extracted from different Bragg reflections indicating departure from thermal equilibrium.

\section{Results and Discussion}

\subsection{Static diffraction and differential scattering signal}

\Cref{fig:aluminium_ediff}b shows the electron diffraction pattern of the polycrystalline aluminium film measured at negative pump-probe delays. We normalized signals to the primary unscattered electron beam (blue circles in \Cref{fig:aluminium_ediff}b), improving the signal-to-noise ratio by up to one order of magnitude. Previous UED studies \cite{Liang2009,Waldecker2015} normalized to the (111) reflection because the primary beam had to be blocked to prevent saturation and damage to scintillator-based detectors, reducing the accessible dynamic range and limiting sensitivity to weak diffraction signals associated with low-amplitude lattice dynamics. The radial intensity distribution (\Cref{fig:aluminium_ediff}a) exhibits ten Bragg reflections from (111) to (551). The pronounced intensity at high diffraction orders, despite the low bunch charge of 22 aC, reflects the high dynamic range of the direct electron detector, combined with a high repetition rate and a low transverse emittance (2.4 \si{\radian\cdot\nano\meter}). \Cref{fig:aluminium_ediff}c shows the two-dimensional map of the pump-probe differential signal $\Delta I/I_0$ as a function of momentum transfer and delay time. Negative values (blue) correspond to suppression of Bragg peak intensity due to increased atomic displacements, while positive values (red) indicate enhanced diffuse scattering between Bragg scattering regions.

\subsection{Bragg scattering dynamics}
\begin{figure*}
	\includegraphics[width=0.8\textwidth]{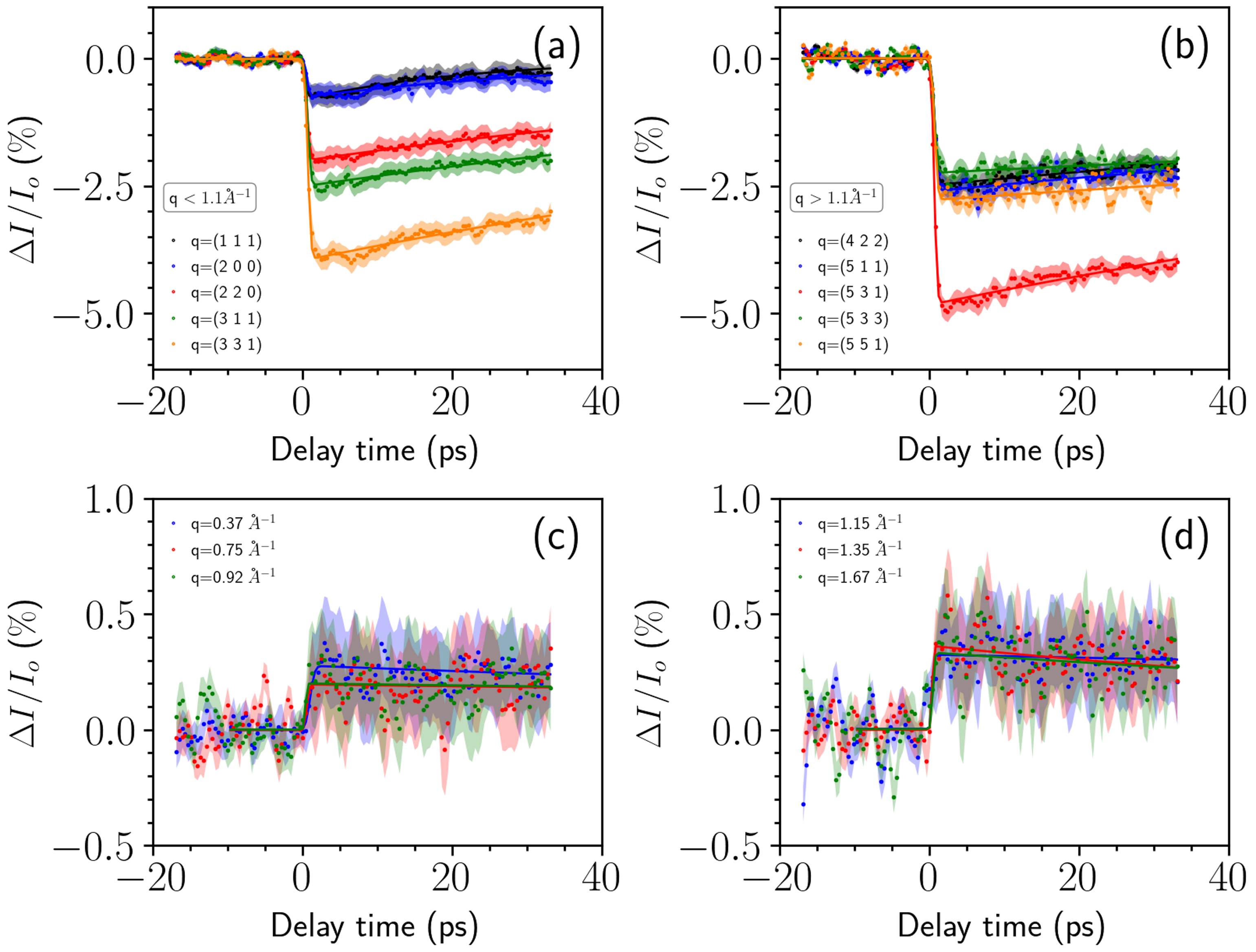}
	\caption{(a-d) Measured $\Delta I/I_0$ of the Bragg (a,b) and diffuse (c,d) scattering signals as a function of pump-probe delay in a photoexcited polycrystalline aluminium thin film for low (a,c) and high (b,d) reciprocal-space, $q$, regions ($q < 1.1$ \si{\angstrom^{-1}} and $q > 1.1$ \si{\angstrom^{-1}}, respectively). Shaded areas indicate error bars. 
	\label{fig:qtraces}}
\end{figure*}

\Cref{fig:qtraces}(a,b) show the time-dependent \(\Delta I/I_0\) signals for ten Bragg reflections separated into long (\(q < 1.1 \, \text{\AA}^{-1}\)) and short (\(q > 1.1 \, \text{\AA}^{-1}\)) range dynamics of reciprocal space regions. All reflections exhibit a rapid decrease in intensity within the first 1--2 picoseconds after photoexcitation, due to energy transfer mechanisms in an electron-phonon coupling framework. Following this initial response, two distinct behaviors emerge: lower-order reflections, such as (111) and (200), relax substantially over tens of picoseconds, whereas higher-order reflections, such as (533) and (551), exhibit significantly slower relaxation or near-plateau behavior.

\begin{table}
	\caption{Measured time constants for the decay and rise of the Bragg and diffuse differential signals, respectively, at specific points in reciprocal space, $q$. \label{tab:aluminium_timeconstants}}
    \begin{ruledtabular}
    \begin{tabular}{@{\extracolsep{\fill}}cccc}
		\multicolumn{2}{c}{\textbf{Bragg scattering regions}} & \multicolumn{2}{c}{\textbf{Diffuse scattering regions}} \\\hline
		$q$ (\si{\angstrom^{-1}}) & Decay time (ps) & $q$ (\si{\angstrom^{-1}})& Rise time (ps) \\\hline
		$\mathrm{q}_{(1 1 1)}$ = 0.428 & 0.918 $\pm$ 0.083 & $\rm{q_{DS-a}}$ = 0.37 & 1.940 $\pm$ 0.300 \\
		$\mathrm{q}_{(2 0 0)}$ = 0.483 & 1.088 $\pm$ 0.113 & \multirow{2}{*}{$\rm{q_{DS-b}}$ = 0.75} &  \multirow{2}{*}{1.018 $\pm$ 0.515}\\
		$\mathrm{q}_{(2 2 0)}$ = 0.693 & 1.152 $\pm$ 0.042  &&\\
		$\mathrm{q}_{(3 1 1)}$ = 0.811 & 1.285 $\pm$ 0.039 & \multirow{2}{*}{$\rm{q_{DS-c}}$ = 0.92} &  \multirow{2}{*}{0.687 $\pm$ 0.308}\\
		$\mathrm{q}_{(3 3 1)}$ = 1.085 & 1.330 $\pm$ 0.033 &&\\
		$\mathrm{q}_{(4 2 2)}$ = 1.203 & 1.373 $\pm$ 0.056 & \multirow{2}{*}{$\rm{q_{DS-d}}$ = 1.15} &  \multirow{2}{*}{0.704 $\pm$ 0.220}\\
		$\mathrm{q}_{(5 1 1)}$ = 1.276 & 1.404 $\pm$ 0.069 &&\\
		$\mathrm{q}_{(5 3 1)}$ = 1.458 & 1.240 $\pm$ 0.033 & \multirow{2}{*}{$\rm{q_{DS-e}}$ = 1.35} &  \multirow{2}{*}{0.695 $\pm$ 0.174}\\
		$\mathrm{q}_{(5 3 3)}$ = 1.623 & 1.293 $\pm$ 0.076 &&\\ 
		$\mathrm{q}_{(5 5 1)}$ = 1.759 & 1.371 $\pm$ 0.079 & $\rm{q_{DS-f}}$ = 1.67 & 0.860 $\pm$ 0.371 \\
    \end{tabular}
    \end{ruledtabular}
\end{table}

The extracted decay times (see \Cref{tab:aluminium_timeconstants}) reveal two characteristic timescales: the three lowest-order reflections (111), (200), and (220) exhibit a mean decay time of $\tau_\mathrm{BS-a} = 1.053 \pm 0.079$ ps, while the higher-order reflections show a slower response with $\tau_\mathrm{BS-b} = 1.328 \pm 0.055$ ps. These timescales are consistent with electron-phonon coupling times reported in previous UED and optical reflectivity studies of aluminium \cite{Guo2000,Nie2006,Zhu2013}.

Our approach of analysing each Bragg reflection independently reveals $q$-dependent dynamics that are obscured when signals are averaged across reflections. Waldecker \emph{et al.} reported a faster average decay time of $\tau = 350 \pm 45$ fs after deconvoluting their 150 fs temporal resolution, but this value represents an average over multiple reflections with different intrinsic response times. The $q$-dependent decay times we observe suggest that the lattice response to electronic excitation is not uniform across reciprocal space. Moreover, higher-order reflections, such as (5 5 3) and (5 5 1), exhibit pronounced oscillations, indicating distinct dynamics associated with short- and long-range interactions that persist beyond the initial electron-phonon equilibration. 

\subsection{Diffuse scattering dynamics}

The diffuse scattering signals (see \Cref{fig:qtraces}c,d) exhibit positive $\Delta I/I_0$, reflecting enhanced scattering from increased lattice disorder. The signals rise rapidly within the first few picoseconds and remain elevated without significant decay over the measured time window. Two distinct rise times are observed (see \Cref{tab:aluminium_timeconstants}): the lowest-$q$ diffuse region exhibits a rise time of $\tau_\mathrm{DS-a} = 1.940 \pm 0.300$ ps, while the higher-$q$ regions show faster rise times with a mean of $\tau_\mathrm{DS-b} = 0.793 \pm 0.318$ ps. The faster timescale is consistent with $\tau = 0.7 \pm 0.1$ ps reported by Zhu \emph{et al.} \cite{Zhu2013}. Moreover, the non-thermal lattice model \cite{Waldecker2016} predicts a hierarchy of equilibration rates, with $\tau_\mathrm{LA} < \tau_\mathrm{TA_1} < \tau_\mathrm{TA_2}$. Our observation of slower (faster) rise times at low (high) $q$ is qualitatively consistent with this prediction, though direct branch assignment is not possible from our powder-averaged data. Waldecker \emph{et al.} \cite{Waldecker2016} reported faster thermalization ($\tau = 0.27 \pm 0.02$ ps) at lower fluence (1.3 \si{mJ/cm^2}), but observed increasing thermalization time with absorbed energy density, exceeding 1 ps above 600 \si{J/cm^3}. Our excitation conditions (2.0 \si{mJ/cm^2}, corresponding to approximately 666 \si{J/cm^3}) are consistent with this trend.

The diffuse scattering at a given $q$ region is sensitive to phonon populations and energy dissipation mechanisms in the lattice. The observation of two distinct rise times in the diffuse scattering signal suggests multiple pathways for energy dissipation mediated by phonon distributions with characteristic timescales.  Following the framework established by Zhu \emph{et al.} \cite{Zhu2013}, the faster rise at high $q$ reflects short-range disorder (short-wavelength phonons) that develops rapidly after electronic excitation, while the slower rise at low $q$ is associated with longer-range correlations (long-wavelength phonons) that evolve on longer timescales.  Furthermore, these findings prompt a discussion on whether the phonon-phonon coupling among branches and the anharmonicity occurring at elevated lattice temperatures\cite{Tang2010,Glensk2019} play a critical role in the nonthermal laser-induced lattice disordering.

\subsection{Mean-square displacement dynamics}

\begin{figure}
	\centering
	\includegraphics[width=\linewidth]{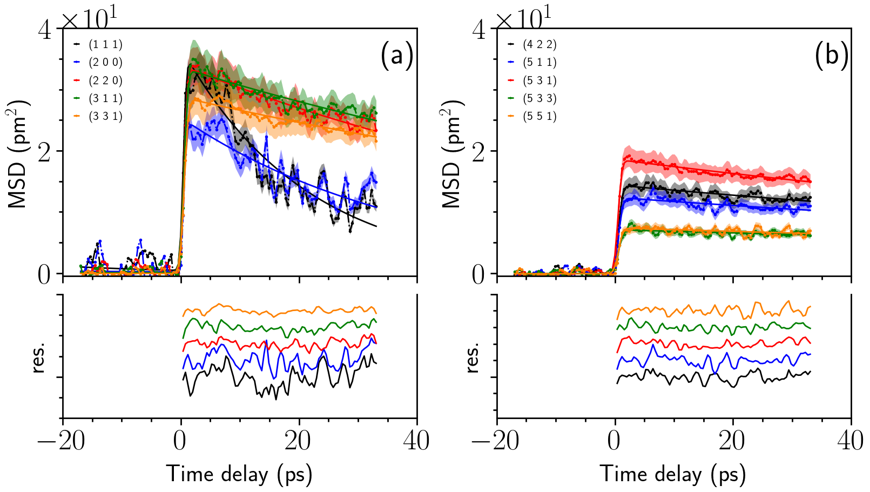}
	\caption{(a-b) Measured mean-square displacement (MSD) extracted from the Bragg scattering signals as a function of pump-probe delay in a photoexcited polycrystalline aluminium thin film for low (a) and high (b) q-space regions ($q < 1.1$ \si{\angstrom^{-1}} and $q > 1.1$ \si{\angstrom^{-1}}, respectively). Shaded areas indicate error bars. \label{fig:aluminium_pumpprobe_msd_traces}}
\end{figure}

To quantify the transient atomic displacements, we extracted the MSD from each Bragg reflection independently using \Cref{eq:MSD}. \Cref{fig:aluminium_pumpprobe_msd_traces} shows the resulting MSD dynamics for low and $q$-space Bragg reflections. All reflections exhibit a prompt increase in MSD at $T_0$, reflecting rapid energy transfer from the excited electronic system to the lattice. However, the subsequent evolution differs markedly between reflections. Low-$q$ reflections such as (111) and (200) display fast MSD relaxation within tens of picoseconds, indicating rapid recovery of the lattice distortion. In contrast, higher-$q$ reflections exhibit slower relaxation or near-plateau behavior, with the MSD remaining elevated throughout the measurement window. This $q$-dependent dynamics, indicating different relaxation mechanisms in the non-thermal regime of lattice disorder, is the central observation of this work. In a thermally equilibrated lattice, the MSD extracted from any Bragg reflection should be identical, since thermal motion is isotropic and the Debye-Waller factor depends only on $\langle u^2 \rangle$ and $q$. The fact that we extract markedly different MSD dynamics from different Bragg orders in reciprocal space demonstrates that the transient lattice state cannot be described by a single thermal distribution. The $q$-dependent MSD recovery is qualitatively consistent with the non-thermal lattice model \cite{Waldecker2016} and with the $q$-dependent rise times observed in the diffuse scattering.

\subsection{Coherent oscillations and frequency analysis}
\begin{figure*}
	\centering
	\includegraphics[width=0.95\textwidth]{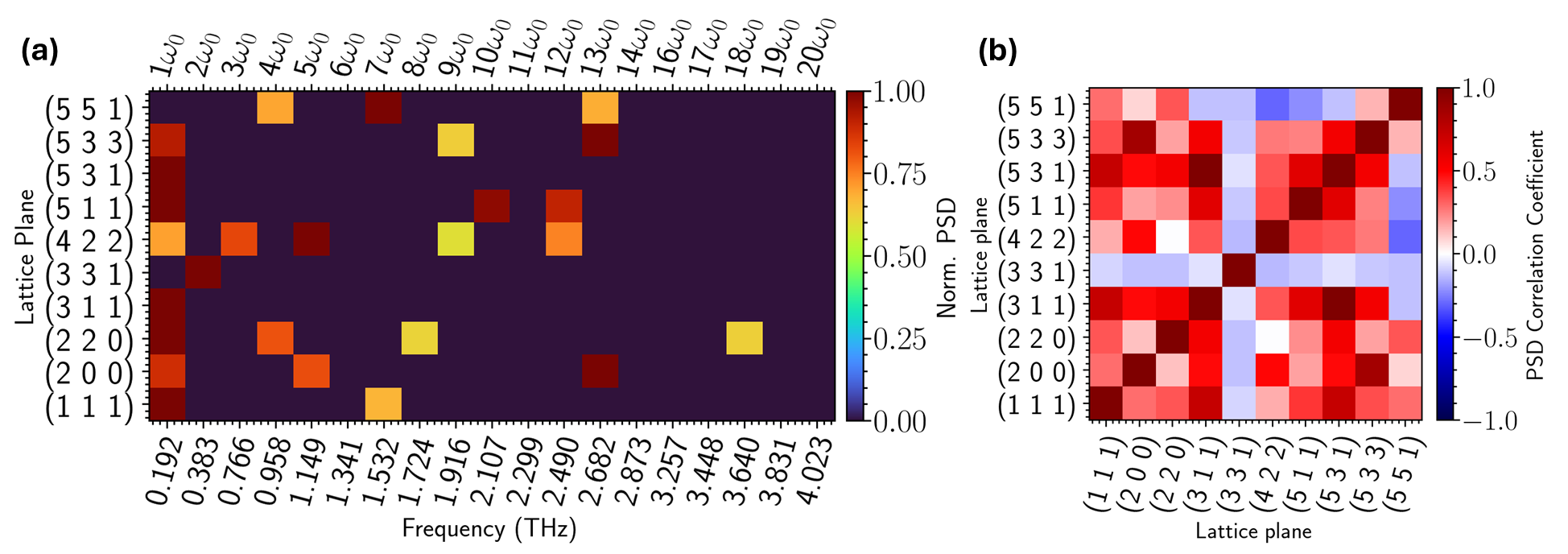}
	\caption{(a) Power spectral density (PSD) of the mean-square displacement fit residuals for the analysed Bragg reflections. (b) Spearman correlation coefficient between the power spectral densities of the mean-square displacement fit residuals for different lattice planes.
	\label{fig:aluminium_pumpprobe_PSD}}
\end{figure*}

Examination of the MSD residuals after subtracting the fitted response reveals oscillatory components that persist well beyond the initial electron-phonon equilibration. \Cref{fig:aluminium_pumpprobe_PSD}a shows the power spectral density (PSD) calculated from the MSD residuals for each Bragg reflection. A fundamental frequency of $\omega_0 = 0.192$ THz is observed across multiple reflections, corresponding to a coherent acoustic breathing mode associated with longitudinal expansion and contraction of the aluminium film. 
Notably, the observation of higher harmonics of \( \omega_0 \) is consistent with the theoretical description of Kuznetsov and Stanton\cite{Kuznetsov1994}, in which coherent phonon oscillations originate from a single zero-wavevector mode occupied by multiple coherent phonons. The oscillation period is given by $\mathcal{T} = 2L/v$, where $L$ is the film thickness and $v$ is the longitudinal sound velocity. Using the room-temperature value $v = 6420$ m/s for aluminium and accounting for the laser-induced temperature rise, which reduces the sound velocity, we estimate an effective value of $v \approx 5952$ m/s, corresponding to a transient lattice temperature increase of approximately 300 K. This estimate is consistent with prior measurements \cite{Waldecker2016}. The (311) reflection has previously been reported to exhibit a breathing mode with period $6.4 \pm 0.5$ ps, corresponding to a frequency of 0.16--0.17 THz \cite{Park2005,Nie2006,Liang2009}, in reasonable agreement with our observed $\omega_0$ (0.192 THz).

To examine correlations between the oscillatory responses across different reflections, we computed the Spearman correlation coefficient between their PSDs, as shown in \Cref{fig:aluminium_pumpprobe_PSD}b. Several features are notable. Firstly, in contrast to the (3 1 1) and (5 3 1) planes, that are strongly correlated and have the same dominant frequency response at \(\omega_0\), the (3 3 1) and (5 5 1) lattice planes, exhibit negligible correlation with other lattice planes and show a single dominant frequency component at \(2\omega_0\) and \(7\omega_0\) (see \Cref{fig:aluminium_pumpprobe_PSD}a), respectively. This may suggest that such planes oscillate, on average, out of phase with the rest of the planes, as in a disordered phase. Additional correlations between higher-order reflections, such as (551), and lower-order reflections, such as (200) and (220), along with the delayed onset of oscillatory amplitudes in the (551) MSD signal, suggest that continued redistribution of vibrational energy among harmonic components can occur at later pump-probe delays.

While branch-resolved phonon identification is not possible in our polycrystalline UED study, the reflection-dependent oscillatory behavior and correlation structure indicate that the coherent lattice response is not uniform across reciprocal space. This is consistent with continued energy redistribution within the phonon system at pump-probe delays well beyond the initial electron-phonon equilibration, as expected for a non-thermal lattice state.

\section{Summary}
We have demonstrated that analysing ultrafast electron diffraction signals separately for each Bragg reflection, rather than averaging across reflections, reveals $q$-dependent lattice dynamics in photoexcited aluminium. The presence of two distinct rise times in the diffuse scattering signal indicates multiple energy-dissipation pathways, each mediated by phonon distributions with characteristic timescales. Coherent oscillations at the breathing mode frequency and its harmonics show reflection-dependent behavior, indicating continued energy redistribution within the lattice at delays well beyond the initial electron-phonon equilibration, as expected for a non-thermal lattice state. While several diffraction planes share the same fundamental oscillation frequency, the spectral content and temporal evolution of the oscillatory response vary across reciprocal space, as evidenced by the PSD and correlation analyses. Overall, the combined time- and frequency-domain analysis provides indirect experimental evidence for mode-selective lattice dynamics and multi-stage energy redistribution in photoexcited aluminium. Although phonon-phonon interactions are not directly resolved, the observed reciprocal-space-dependent relaxation dynamics, delayed MSD recovery at high scattering vectors, and evolution of coherent oscillatory components are consistent with a non-thermal lattice description in which energy is redistributed among multiple phonon populations on different timescales. These results demonstrate that high-sensitivity UED measurements, with single-electron detection and appropriate signal normalization, can reveal non-equilibrium dynamics that are obscured in conventional diffraction-averaged analyses.

\begin{acknowledgments}
We acknowledge financial support from the European Research Council for ERC Starting Grant ``TERES'' (Grant No. 101165245). We are grateful to Mark Mero, Christoph Reiter, Arnaud Rouz\`ee, Roman Peslin, Wolfgang Krueger, Thomas Mueller, and Marc J. J. Vrakking for support. 
\end{acknowledgments}

\bibliographystyle{apsrev4-1}
\bibliography{aipsamp}

\end{document}